\documentstyle[12pt,psfig,preprint]{aastex}

\def\pbar{\overline{p}}
\def\tpbar{\tau_{\overline{p}}}

\font\eightrm=cmr10 scaled 800

\slugcomment{}
\shorttitle{New Limit on Antiproton Lifetime}
\shortauthors{Geer \& Kennedy}

\begin{document}

\title{A NEW LIMIT ON THE ANTIPROTON LIFETIME}
\author{{\bf Stephen H. Geer}}
\authoraddr{Fermi National Accelerator Laboratory}
\affil{Fermi National Accelerator Laboratory, P.O. Box 500, Batavia, 
Illinois 60510}
\email{sgeer@fnal.gov}
\and
\author{{\bf Dallas C. Kennedy}}
\authoraddr{University of Florida}
\affil{Department of Physics, University of Florida, Box 118440, Gainesville FL 32611}
\email{kennedy@phys.ufl.edu}
\affil{\rm FERMILAB-PUB-98/383-A $\bf{\bullet}$ UF-IFT-HEP-98-34\\ 
Original: December 1998 $\bf{\bullet}$ Revised: July 1999}
\begin{abstract}
Measurements of the cosmic ray $\pbar /p$ ratio are 
compared to predictions from an inhomogeneous disk-diffusion model 
of $\pbar$ production and propagation within the Galaxy, 
combined with a calculation of the modulation of the interstellar 
cosmic ray spectra as the particles propagate through the heliosphere 
to the Earth. The predictions agree with the observed $\pbar /p$ 
spectrum. Adding a finite $\pbar$ lifetime to the model, we obtain 
the limit $\tpbar >$ 0.8 Myr (90\% C.L.).
\end{abstract}

\indent\centerline{\hskip-40pt Published in {\bf the Astrophysical Journal}:
20 March 2000}

\section{Introduction}

In recent years the presence of antiprotons $(\pbar$'s) in the 
cosmic ray (CR) flux incident upon the Earth has been firmly 
established by a series of balloon--borne experiments
(Golden et al. 1979; Bogomolov et al. 1979; Bogomolov et al. 1987; Bogomolov
et al. 1990; Hof et al. 1996; Mitchell et al. 1996; Moiseev et al. 1997;
Yoshimura et al. 1995; Boezio et al. 1997; Matsunaga et al. 1998).
The measurements are summarized in Table~1. 
The observed CR $\pbar /p$ ratio has been shown to be in approximate
agreement with predictions based on the Leaky Box Model (LBM)
(Stephens 1981; Stephens \& Golden 1987; Webber \& Potgieter 1989; 
Gaisser \& Schaefer 1992),
which assumes that the $\pbar$'s originate from proton interactions in the 
interstellar medium (ISM). The $\pbar$'s then propagate within the 
Galaxy until they ``leak out'' by diffusion and convection
with the characteristic CR Galactic escape time $T\sim 10$~million years (Myr) 
(Webber \& Potgieter 1989; Gaisser \& Schaefer 1992).  
If the $\pbar$ lifetime $\tpbar$ is not long compared to $T$ the 
predicted $\pbar /p$ spectrum will be modified. 
The agreement of the LBM predictions with the observed 
$\pbar /p$ spectrum has therefore been used to argue that $\tpbar > 10$~Myr
(Golden et al. 1979; Bogomolov et al. 1979; Stephens 1981; Stephens \& 
Golden 1987).
This estimated limit is based on early CR $\pbar$ data and does not take into 
account the reduction of the $\pbar$ decay rate due to time dilation, 
the effect of the heliosphere 
on the observed $\pbar /p$ spectrum, 
or the systematic uncertainties associated with the predictions. In this 
paper we compare recent CR data
with the predictions of an improved LBM extended to permit a finite $\tpbar$. 
Heliospheric corrections and 
systematic uncertainties are taken into account. Assuming a stable $\pbar$, 
we find excellent agreement between our predictions and the CR observations. 
Allowing the $\pbar$ to decay, we obtain a lower 
limit on $\tpbar$ which is 
significantly more stringent than current laboratory bounds 
obtained from searches for $\pbar$ decay in ion
traps (Gabrielse et al. 1996)
and storage rings (Geer et al. 1994; Hu et al. 1998).
The analysis presented in this paper improves on our earlier 
analysis (Geer \& Kennedy 1998)
by including new data from Matsunaga et al. (1998).

CPT invariance requires $\tpbar$ = $\tau_p$, where the proton lifetime 
$\tau_p$ is known to exceed ${\cal O}(10^{32})$ yr (Caso et al. 1998).
Although there is no compelling theoretical motivation to suspect a violation 
of CPT invariance, and hence a short $\pbar$ lifetime, it should be noted that 
string theories can accommodate CPT violation.  Consider a mass--dimension--$n$
CPT--violating quantum field operator suppressed by the characteristic scale 
$m_X$, with $n>$ 4.  Dimensional analysis provides the estimate $m_p\tpbar\sim 
[m_p/m_X]^{2n-8}$, yielding $m_X/m_p\sim [4.5\times 10^{38}\cdot
\tpbar /10{\rm\ Myr}]^{1/(2n-8)}$ (Kennedy 1999).  For a given lower limit
on $\tpbar$, the implied lower limit on $m_X$ is most stringent for $n$ = 5.
Note that if $m_X$ is at the Planck scale $(1.2\times 10^{19}$ GeV/$c^2)$ and 
$n$ = 5, the expected $\tpbar$ would be $\sim$ 10 Myr.
Hence, a search for $\pbar$ decay with a lifetime approaching 10~Myr 
provides a test for CPT violation well beyond the scale accessible at 
high energy colliders. 
Finally, since the antiproton is the only long lived antiparticle 
that could decay into other known particles without 
violating charge conservation, a search for a modification of the CR 
$\pbar$ spectrum by $\pbar$ decay provides a unique test of the 
stability of antimatter.

\section{Galactic Production and Propagation}

In the LBM the ISM $\pbar$'s are produced by the interactions of 
CR $p$'s (Stephens 1981; Stephens \& Golden 1987; Webber \& Potgieter 1989;
Gaisser \& Schaefer 1992):
$pN_Z\rightarrow\pbar X,$ where $N_Z$ is a nucleus of charge $Z$, 
and $X$ is anything. Our calculations use the elemental ISM abundances 
given in Webber \& Potgieter (1989) and Gaisser \& Schaefer (1992),
and the measured cross sections for $Z$ = 1 (the dominant contribution) 
given in Stephens (1981).
For $Z >$ 1, we have used the ``wounded nucleon'' results also used by Gaisser
\& Schaefer (1992).

The $\pbar$'s are assumed to propagate within the Galaxy until they are lost 
either by leakage into intergalactic space or by $p\pbar$ annihilation.  The
dominant loss process is leakage.  Energy loss is included in the model.
Comparisons between predicted and observed CR isotopic abundances
imply that reacceleration can be neglected (Webber et al. 1992).  
Reacceleration is not included in our calculation.  The LBM incorporates
space-dependent diffusive and convective CR leakage into a single escape term
with a single characteristic escape time.  This picture is known to be too
simple.  A more accurate picture of Galactic CRs (Webber et al. 1992) is
provided by a two-zone diffusion-convection model which we refer to as the
{\it inhomogeneous leaky disk model} (ILDM).  The model is based on a $h\simeq$
100 pc-thick matter disk (radius $R\simeq$ 20 kpc), with a halo extending above
and below the disk by $L\simeq\pm 3$ kpc.  The CRs diffuse off the disk
into the halo by magnetic turbulence.  It has been shown that convection driven
by the Galactic wind can be neglected (Bloemen 1993).  The ILDM {\it disk} 
storage time $\sim L^2/4{\cal K},$ where the diffusion transverse to the
Galactic disk ${\cal K}$ $\simeq (6\pm 4)\times 10^{23}$ m$^2$ sec$^{-1}$
(Bloemen 1989; Webber et al. 1992).  
The ILDM with these parameters $(h,R,L,{\cal K})$ has been numerically solved 
by Chardonnet et al. (1996) for the secondary CR $\pbar$ spectrum.  Although
the ILDM provides a more realistic physical description of CR propagation than
the LBM, Chardonnet et al. found that the fluxes in the Galactic disk predicted
by the ILDM are reproduced within uncertainties by a modified LBM (MLBM)
in which the
momentum-dependent escape time $T(P)$ is given by: $T(P) = $~(13~Myr)~$[1+
cP/(3{\rm\ GeV})]^{-0.6}$.  This escape time is longer than that used by Webber
\& Potgieter (1989)
and Gaisser \& Schaefer (1992) and, since the $\pbar$ flux is proportional to
$T(P)$, raises the predicted $\pbar$ flux and continues the trend of the last
two decades toward longer CR escape times/lengths, as discussed by these 
authors.

The uncertainties on the parameters of the MLBM result in uncertainties 
on the normalization of the predicted $\pbar /p$ ratio but, to a good 
approximation, do not introduce significant uncertainties in the shape 
of the predicted spectrum. Uncertainties on four parameters must be 
considered: (i) the storage time ($\pm$67\% ) (Webber et al. 1992),
(ii) the ISM primary $p$ flux ($\pm$35\% ) (Gaisser \& Schaefer 1992),
(iii) the $\pbar$ production cross section ($\pm$10\% ) (Stephens 1981;
Gaisser \& Schaefer 1992),
and (iv) the composition of the ISM, which introduces an uncertainty of 
$<6$\% on the predicted $\pbar$ flux (Gaisser \& Schaefer 1992).
We neglect the last of these uncertainties since it is relatively small. 
Within the quoted fractional uncertainties on the other three 
parameters we treat all values as being a priori equally likely. 
Note that the predicted $\pbar /p$ ratio is approximately proportional 
to each of the parameters under consideration. 

\section{Heliospheric Modulation}

The solid curves in Figure~1 show the MLBM $\pbar /p$ 
spectra for the parameter choices that result in the largest and smallest 
$\pbar /p$ predictions. The predicted ISM spectrum does not give a good
description of the observed distribution at the top of the atmosphere.
Good agreement is not expected because
the CR spectra observed at the Earth are modulated as the particles propagate 
through the heliosphere, which consists of the solar magnetic field ${\bf B}$ 
and solar wind (Stix 1989; Encrenaz et al. 1990; Longair 1992).
The wind, assumed to blow radially outwards, has equatorial speed 
$V_W\simeq$~400~km~sec$^{-1}$.  {\it Ulysses} has found a latitude-dependent 
$V_W(\theta )$.  At high latitudes $(|\theta -90^\circ |\gtrsim 20^\circ )$,
$V_W\simeq$~750~km~sec$^{-1}$ 
(Smith et al. 1995; Marsden et al. 1996; Ulysses 1999).  
The latitude-dependent solar rotation $\Omega_\odot(\theta )$ (Howard 1984) 
twists the field lines into a Parker spiral. The smoothed heliomagnetic field 
($B_\oplus\sim$ 5 nT at the Earth's orbit) declines as it
changes from radial at the Sun to azimuthal in the outer solar system.  
The heliomagnetic polarity ${\rm sgn}(A)$ is opposite in northern and 
southern solar hemispheres and switches sign somewhat after sunspot maximum.
Note that $A>$ 0 in the 1990-99 epoch.
The regions of opposite magnetic polarity are separated by an
approximately equatorial, unstable neutral current sheet. 
The sheet is spiraled and wavy; its waviness is measured by the tilt angle 
$\alpha$, which relaxes from $\simeq$ $50^\circ$ at polarity reversal to 
$\simeq$ $10^\circ$ just before reversal.  The wavy sheet locus in 
heliographic co-ordinates is $\cos\theta + \sin\alpha\cdot\sin 
[\varphi + \Omega_\odot r/V_W]$ = 0 (Jokipii \& Kopriva 1979;
Encrenaz et al. 1990; NSO 1999).

Cosmic rays enter the heliosphere on ballistic trajectories and are then
transported by drift and diffusion
(Jokipii \& Kopriva 1979; Risken 1984).  The transport
model includes energy loss as CRs perform work against the wind, CR solar 
wind convection, drift of CRs across inhomogeneous field
lines, and diffusive scattering by field turbulence.  Particles with $qA > 0$ 
$(qA < 0)$ drift in along a polar (sheet) route.  Since the polar $V_W$ is
approximately twice the equatorial $V_W$, particles with $qA>$ 0 find it
more difficult to drift in along a polar route than do $qA<$ 0 particles along
a sheet route.  In the following, we restrict our analysis
to CR $p$'s and $\pbar$'s with kinetic energies $>$ 500 MeV.  At these energies,
(a) the wind and magnetic drift (Isenberg \& Jokipii 1979, 1981) dominate
the transport; and (b) the diffusion is charge--symmetric and, for a 
given rigidity,
modulates the $p$'s and $\pbar$'s by the same factor, and hence cancels in the
ratio of fluxes (see Ulysses $e$ and $p$ scans: Smith et al. 1995;
Marsden et al. 1996; Fisk et al. 1998).

We compute the modulation of the CR fluxes by the method of characteristics,
numerically implemented in three dimensions by a combination of Runge--Kutta 
and Richardson--Burlich--Stoer techniques (Press et al. 1992).  
The calculation is based on the heliospheric transport
model of Jokipii \& Kopriva (1979), Jokipii \& Thomas (1981), and
Jokipii \& Davila (1981), but that model has been updated with {\it Pioneer,
Voyager, Helios, IMP} and {\it Ulysses} heliospheric measurements (Smith et al.
1995; Marsden et al. 1996; NSSDC 1999; SEC 1999).  The improvements include
the latitude-dependent $V_W(\theta )$ and $\Omega_\odot(\theta )$.
The calculation is simplified by ignoring diffusion where its effects on the
$\pbar /p$ ratio are small, which for our energy range 
means everywhere except across the sheet (Jokipii \& Thomas 1981).
The effects of diffusion away from the sheet are essential for particles with
energies below the range we are considering.  For particles with speed $v$
the across-sheet diffusion coefficient $\kappa_\perp$ = 
$[(2-3)\times 10^{17}$~m$^2$/sec$][B_\oplus /B(r)](P/{\rm GeV})^{1/3}(v/c)$ is 
used (Smith et al. 1995; Marsden et al. 1996).  The implied field fluctuation
spectral index is the Kolmogorov value of 5/3.
The modulated fluxes at Earth depend only weakly on $\kappa_\perp$
in our energy range, for this and larger values of $\kappa_\perp$.

\section{Results}

We use the data sets
recorded by the MASS91, IMAX, BESS, and CAPRICE experiments (Table~1). 
These data were recorded in the period 1991--1995, corresponding to a 
well--behaved part of the solar cycle for which the heliospheric modulation 
corrections can be confidently calculated. 
To explore the dependence of the predicted spectrum on the heliospheric 
parameters (equatorial $V_W$, polar $V_W$, and $B_\oplus$) we have computed 
the modulated spectra for 11 parameter sets (F1 -- F11) that span the 
range of acceptable parameter values (Table~2). 
Using the central parameter values for our MLBM, and assuming a stable 
antiproton,  
the predicted modulated $\pbar /p$ spectra for a fixed time in the 
solar cycle (July 1995) are shown in Figure~2 for each of the 11 
heliospheric parameter sets.  Figure~2 also shows the epoch--corrected 
measured CR spectra, obtained by multiplying each measurement by the 
factor $f \equiv R($July 1995$)/R(t)$, where $R(t)$ is the predicted 
$\pbar /p$ ratio at time $t$. The factors $f$, which are shown in Table~1 
and have been computed using 
the F6 parameters, vary by up to $\pm 0.06$ with the parameter set choice. 
The MLBM predicted $\pbar /p$ spectra give an excellent description of the 
measurements.
There is no evidence for an unstable $\pbar$. 

To obtain a limit on $\tpbar$ we add to the MLBM one additional loss
mechanism, $\pbar$ decay.  The results from maximum likelihood fits to the 
measurements are shown in Figure~3  
as a function of the assumed $\tpbar$ for the 11 heliospheric parameter sets. 
The fits, which take account of the Poisson
statistical fluctuations on the number of observed events and the
background subtraction for each data set (Table~1), also 
allow the normalization of the MLBM predictions to vary within 
the acceptable range (Figure~1). 
For a stable $\pbar$, at the 95\% C.L. all of the heliospheric parameter 
sets yield predictions that give reasonable descriptions of the observed 
$\pbar /p$ spectrum. Allowing for a finite $\tpbar$, the heliospheric 
parameter sets with larger wind speeds permit lower $\pbar$ lifetimes. 
This can be understood by noting that, as the wind speed increases, 
the predicted flux of polar--routed particles (protons in the present solar 
cycle) is depleted at kinetic energies $\lesssim$ 10 GeV, which increases 
the predicted $\pbar /p$ ratio in this energy range.  Antiproton decay would 
compensate for this distortion in the predicted spectrum. 
Hence our fits using the extreme parameter set F11 determine the limits 
on $\tpbar$.  Under the assumption that there are no significant non--standard 
sources of cosmic ray antiprotons, we obtain the bounds:
\begin{equation}
\tpbar\quad >\quad 0.8\ {\rm Myr\ \ (90\%\ C.L.)}\quad ,\quad
                   0.7\ {\rm Myr\ \ (95\%\ C.L.)}\quad ,\quad
                   0.5\ {\rm Myr\ \ (99\%\ C.L.)}\quad .\nonumber \\
\end{equation}
Our simple dimensional analysis suggests that if a dimension--$n$ 
CPT--violating coupling results in antiproton decay, the mass scale $m_X$ at 
which this new physics takes place exceeds ${\cal O}(10^{19})$~GeV/$c^2$
$[{\cal O}(10^9)$~GeV/$c^2]$, for $n$ = 5~(6).

The limits~(1) are significantly more stringent than those obtained from 
the most sensitive laboratory search for inclusive $\pbar$ decay 
($\tpbar > 3.4$~months) (Gabrielse et al. 1996)
or the most sensitive search for an exclusive $\pbar$ decay mode 
($\tpbar/B(\pbar \rightarrow \mu^-\gamma) > 0.05$~Myr) (Hu et al. 1998).  Note
however that these limits are less restrictive than the estimate $\tpbar$
$\gtrsim$ 10 Myr given in Caso et al. (1998), which takes no account of time
dilation, systematic uncertainties, or propagation effects and which is
therefore overconstraining.

\acknowledgments

It is a pleasure to thank J.~R.~Jokipii (Lunar \& Planetary Laboratory, Univ.
Arizona) and E.~J.~Smith (Jet Propulsion Laboratory/CalTech) for their 
insights.
This work was supported at Fermilab under grants U.S. DOE DE-AC02-76CH03000 
and NASA NAG5-2788 and at the Univ. Florida, Institute for Fundamental Theory, 
under grant U.S. DOE DE-FG05-86ER40272.

\clearpage
\thispagestyle{plain}
\textheight 8.5in

\begin{table}
\begin{center}
\caption{Summary of Cosmic Ray Antiproton Results}
\eightrm
\begin{tabular}{|lc|c|c|c|c|c|c|c|c|}
\hline
Experiment& &Field&Flight&$f^{\rm b}$&KE Range&Cand-&Back-&Observed 
            &Predict-\\
            & &Pol.$^{\rm a}$&Date&&(GeV)&idates&ground&$\overline{p}/p$ Ratio&
            tion$^{\rm c}$\\
\hline
Golden et al. 1979$^\dag$&&$+$&June 1979&--&5.6 -- 12.5&46&18.3&
$(5.2\pm 1.5)\times 10^{-4}$& -- \\ 
\hline
Bogomolov et al. 1979$^\dag$&&$+$&1972-1977&--&2.0 -- 5.0&2&--&
$(6\pm 4)\times 10^{-4}$& -- \\
Bogomolov et al. 1987$^\ddag$&&$-$&1984-1985&--&0.2 -- 2.0&1&--&
$(6^{+14}_{-5})\times 10^{-5}$& -- \\
Bogomolov et al. 1990$^\ddag$&&$-$&1986-1988&--&2.0 -- 5.0&3&--&
$(2.4^{+2.4}_{-1.3})\times 10^{-4}$& -- \\ \hline
MASS91$^{\rm d}$&&$+$&Sep. 1991&1.1&3.70--19.08&11&3.3&
$(1.24^{+0.68}_{-0.51})\times 10^{-4}$&$1.3\times 10^{-4}$\\
\hline
IMAX$^{{\rm e}\ddag}$&&$+$&July 1992&--&0.25 -- 1.0&3&0.3&
$(3.14^{+3.4}_{-1.9})\times 10^{-5}$&$1.5\times 10^{-5}$\\
IMAX$^{\rm e}$&&$+$&July 1992&0.96&1.0 -- 2.6 &8&1.9&
$(5.36^{+3.5}_{-2.4})\times 10^{-5}$&$6.5\times 10^{-5}$\\
IMAX$^{\rm e}$&&$+$&July 1992&1.1&2.6 -- 3.2 &5&1.2&
$(1.94^{+1.8}_{-1.1})\times 10^{-4}$&$1.1\times 10^{-4}$\\
\hline
BESS93$^{{\rm f}\ddag}$&&$+$&July 1993&--&0.20 -- 0.60&7&$
\sim 1.4$&$(5.2^{+4.4}_{-2.8})\times 10^{-6}$&$8.9\times 10^{-6}$\\
\hline
CAPRICE$^{\rm g}$&&$+$&Aug. 1994&0.94&0.6 -- 2.0 &4&1.5&
$(2.5^{+3.2}_{-1.9})\times 10^{-5}$&$3.5\times 10^{-5}$\\
CAPRICE$^{\rm g}$&&$+$&Aug. 1994&1.0&2.0 -- 3.2 &5&1.3&
$(1.9^{+1.6}_{-1.0})\times 10^{-4}$&$1.1\times 10^{-4}$\\
\hline
BESS95$^{{\rm h}\ddag}$$^\ast$&&$+$&July 1995&1.0&0.175 -- 0.3&3&0.17&
$(7.8^{+8.3}_{-4.8})\times 10^{-6}$&$-$\\
BESS95$^{{\rm h}\ddag}$$^\ast$&&$+$&July 1995&1.0&0.3 -- 0.5&7&0.78&
$(7.4^{+4.7}_{-3.3})\times 10^{-6}$&$1.1\times 10^{-5}$\\
BESS95$^{{\rm h}\ast}$&&$+$&July 1995&1.0&0.5 -- 0.7&7&1.4&
$(7.7^{+5.3}_{-3.7})\times 10^{-6}$&$5.5\times 10^{-6}$\\
BESS95$^{{\rm h}\ast}$&&$+$&July 1995&1.0&0.7 -- 1.0&11&2.8&
$(1.01^{+5.7}_{-4.3})\times 10^{-5}$&$1.3\times 10^{-5}$\\
BESS95$^{{\rm h}\ast}$&&$+$&July 1995&1.0&1.0 -- 1.4&15&3.5&
$(1.99^{+0.91}_{-0.73})\times 10^{-5}$&$3.1\times 10^{-5}$\\
\hline
\end{tabular}
\end{center}
\vspace{0.25cm}
\indent{${}^{\rm a}$~Heliomagnetic field polarity ${\rm sgn}(A)$.}\\
\indent{${}^{\rm b}$~Epoch correction factor; see text.}\\
\indent{${}^{\rm c}$~MLBM prediction using F6 heliospheric parameters; 
see text.}\\
\indent{${}^{\rm d}$~Hof et al. 1996.  ${}^{\rm e}$~Mitchell et al. 1996.  
${}^{\rm f}$~Moiseev et al. 1997.  ${}^{\rm g}$~Boezio et al. 1997.  
${}^{\rm h}$~Matsunaga et al. 1998.}\\
\indent{${}^\dag$~Not shown in Fig.~1 or used in analysis.  ${}^\ddag$~Not 
used in analysis.  ${}^\ast$~Statistical and systematic uncertainties on ratio 
added in quadrature.}
\label{balloon_tab}
\end{table}

\begin{table}
\centerline{Table 2: Heliospheric Parameter Sets}
\begin{center}
\vspace{0.75cm}
\begin{tabular}{|l|ccc|}
\hline
Set&Equatorial $V_W$ & Polar $V_W$ & $B_{\oplus}$\\
   &(km~sec$^{-1}$)&(km~sec$^{-1}$)&(nT)\\
\hline
F1 & 375 & 700 & 4.0\\
F2 & 380 & 710 & 4.1\\
F3 & 385 & 720 & 4.2\\
F4 & 390 & 730 & 4.3\\
F5 & 395 & 740 & 4.4\\
F6 & 400 & 750 & 4.5\\
F7 & 405 & 760 & 4.6\\
F8 & 410 & 770 & 4.7\\
F9 & 415 & 780 & 4.8\\
F10 & 420 & 790 & 4.9\\
F11 & 425 & 800 & 5.0\\
\hline
\end{tabular}
\label{parameter_tab}
\end{center}
\end{table}

\begin{figure*}[tl]
\psfig{file=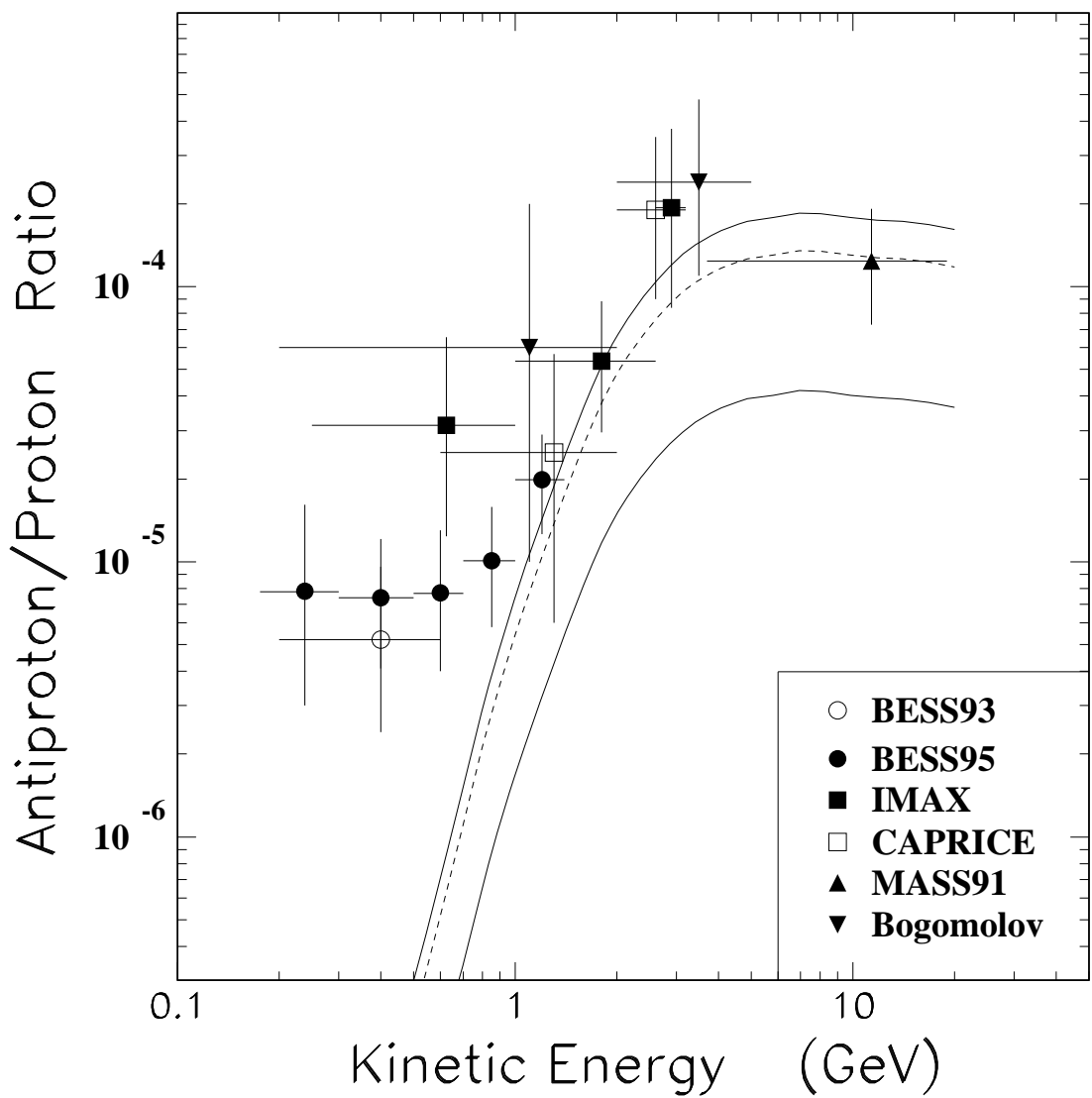,width=6.8in,height=7.2in}
\ \\\
\caption{
Observed $\pbar /p$ ratio at the top of Earth atmosphere (see Table~1).  
The solid curves show the upper and lower interstellar ratios 
predicted by the MLBM described in the text, without solar modulation. 
The broken curve shows the MLBM prediction with the same parameters 
used for the modulated predictions of Fig.~2.}
\label{fig1}
\end{figure*}

\begin{figure*}[tl]
\leavevmode\centering\psfig{file=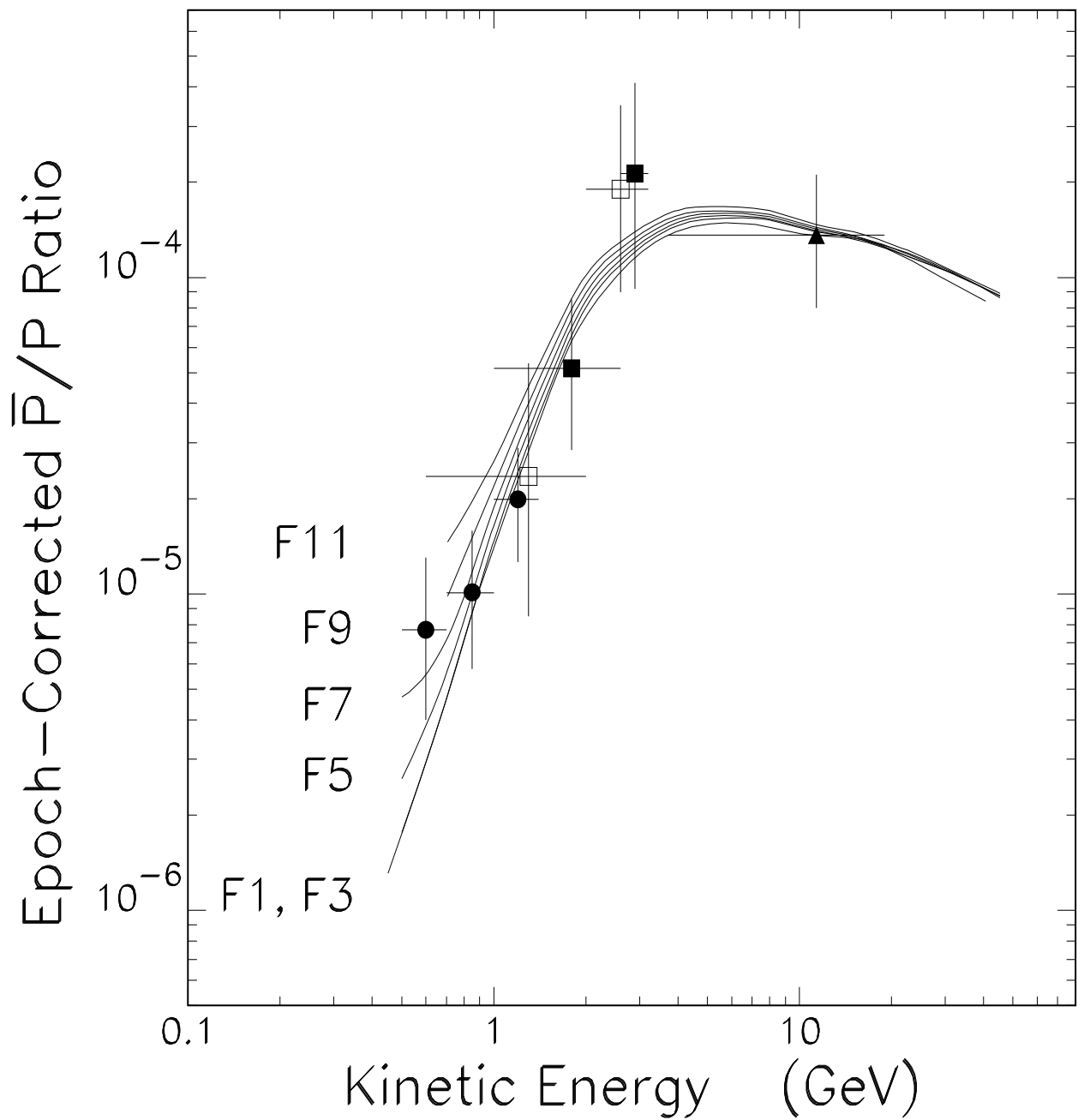,width=6.8in,height=7.2in}
\ \\\ \\\ \\
\caption{
Observed $\pbar /p$ spectrum (kinetic energy $>$ 500~MeV) 
at the top of the atmosphere compared with 
the MLBM predictions (see broken curve on Fig.~1) after 
modulation using the heliospheric parameter sets indicated 
(see Table~2). The curves are predictions for the spectrum observed in 
July 1995. The data have been corrected to correspond to this epoch (see 
text).
}
\label{fig2}
\end{figure*}

\begin{figure*}[tl]
\leavevmode\centering\psfig{file=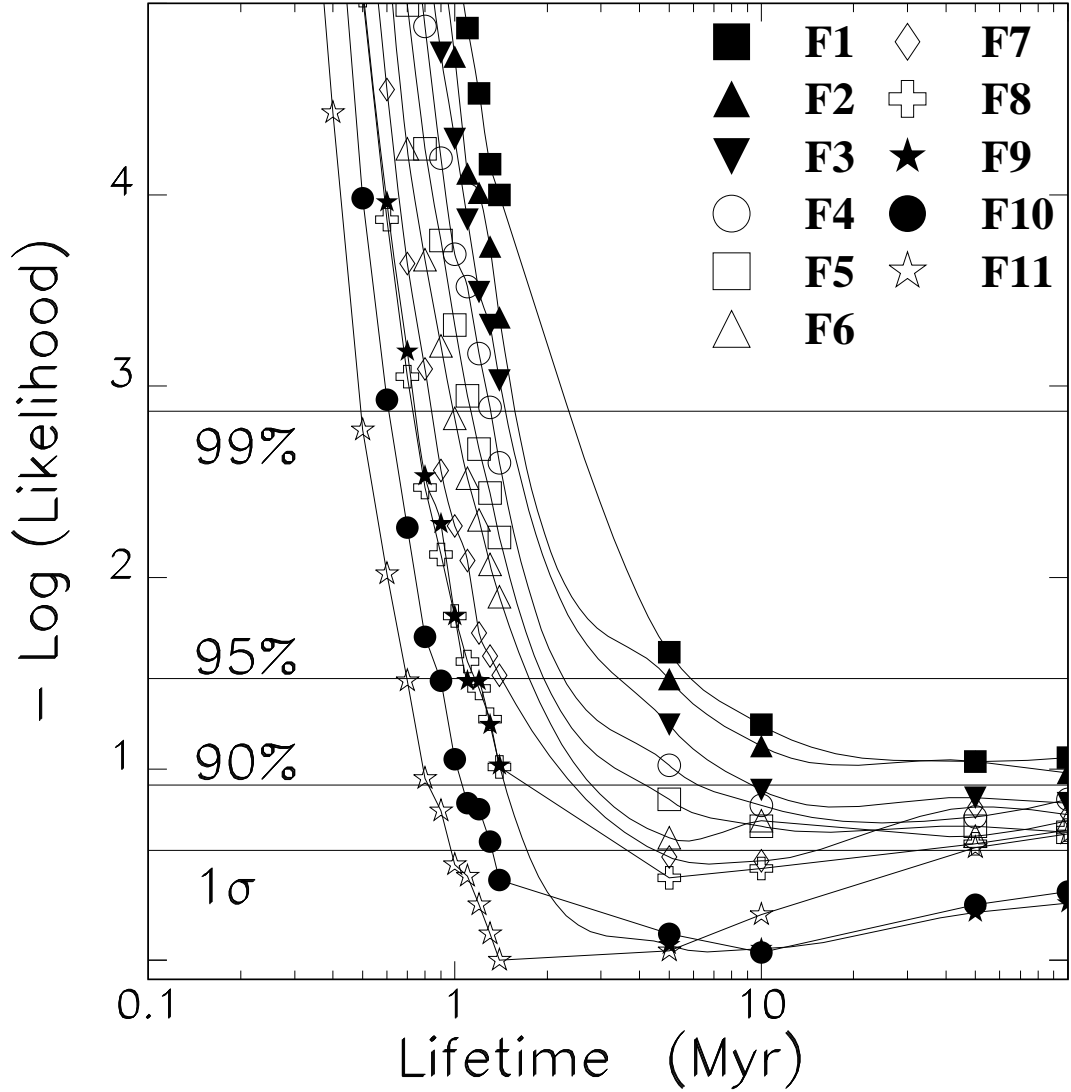,width=6.8in,height=7.2in}
\ \\\
\caption{
Fit results as a function of the assumed $\tpbar$ 
for the eleven heliospheric parameter sets (F1 -- F11) shown in 
Table~2.}
\label{fig3}
\end{figure*}

\end{document}